\documentclass[12pt,a4paper]{article}

\usepackage{indentfirst}
\usepackage[english]{babel}
\usepackage[T1]{fontenc}
\usepackage{textcomp} 
\usepackage[utf8]{inputenc} 
\usepackage{lmodern}
\usepackage{amsmath,exscale,
amssymb,latexsym,varioref,
mathrsfs,a4,textcomp,moreverb,
color,graphicx,lscape,multicol}
\usepackage{setspace}

\begin{document}
 \title{\textbf{Hydrodynamics of a quark droplet}}
  \author{Johan J. Bjerrum-Bohr$^{1,3 }$, Igor N. Mishustin$^{1,2,3}$, Thomas D{\o}ssing$^{3}$\\
 \small $^{1}$ Frankfurt Institute for Advanced Studies (FIAS),\\
 \small Goethe-University, Ruth-Moufang Str. 1, 60438 Frankfurt am Main\\
 \small $^{2}$Kurchatov Institute, Russian Research Center,\\
 \small Akademika Kurchatova Sqr., Moscow, 123182, Russia\\
 \small $^{3}$Niels Bohr Institute, University of Copenhagen,\\
 \small Blegdamsvej 17, 2100 K{\o}benhavn {\O}, Denmark}

\maketitle

\begin{abstract}
We present a simple model of a multi-quark droplet evolution based on the hydrodynamical description. This model includes collective expansion of the droplet, effects of the vacuum pressure and surface tension.  The hadron emission from the droplet is described following Weisskopf's statistical model. We have considered evolution of baryon-free droplets which have different initial temperatures and expansion rates. As a typical trend we observe an oscillating behavior of the droplet radius superimposed with a gradual shrinkage due to the hadron emission. The characteristic life time of droplets with radii 1.5-2 fm are about 9-16 fm/c.
\end{abstract}

\section{Introduction}
In relativistic heavy ion collisions, a hot and dense fireball is produced in the overlap zone. It is believed that at early times this fireball is made of Quark-Gluon Plasma (QGP).  Experiments aimed to study properties of QGP are going on at Brookhaven National Laboratory's Relativistic Heavy Ion Collider (RHIC) and at CERN's new Large Hadron Collider (LHC). Large amounts of data are already accumulated that helps to develop realistic theoretical models of the collision process and deconfinement-hadronization transition. No single approach has so far been able to describe all the aspects of this complicated process, however, several successful models have been proposed. Among the most popular is the hydrodynamical model, where the QGP is assumed to evolve as an almost perfect fluid. Such approach was first proposed by Landau almost 60 years ago \cite{Lan53}, and since that time its different versions are used to describe nuclear collisions at intermediate and high energies (see the recent paper \cite{Mer11} and references therein).

One may expect,  that in the course of fast expansion, this fluid will split into many droplets which later on evolve by evaporating hadrons from the surface \cite{Mis99b,Mis09}. The idea is that the strong collective expansion of matter, as seen in relativistic heavy-ion collisions, may lead to overcooling and subsequent fragmentation of the plasma phase into droplets surrounded by a low-density hadron gas. The dynamical implications of such a scenario were studied in refs. \cite{Mis99a,Cse95, Sca01,Ran04,Mis08}. In this paper, we will present a simple model of relativistic dynamics of a hot quark-antiquark droplet which may be created in this first step of the hadronization of a rapidly-expanding quark-gluon plasma. Our model is motivated by earlier works \cite{Bir83,Bar90}, where a simplified hydrodynamic model with hadron emission was used to describe the dynamics of a quark(-gluon) droplet. In our present approach we additionally consider effects of the surface tension and use the entropy balance equation for determination of the temperature. Also we present a detailed study of the droplet trajectories in collective coordinates.

This paper is organized as follows. In Section 2, starting from hydrodynamic equations, we derive a simple model to describe evolution of an individual quark droplet. In Section 3, semi-analytical solutions of this model are obtained for the idealized case of no hadron emission. In Section 4, the hadron emission from the droplet is included following a statistical approach. In Section 5, we present results of numerical simulations of the droplet evolution taking into account the hadron emission. Conclusion and outlook are given in Section 6.

\section{Modeling evolution of a quark droplet}
\subsection{Averaging hydrodynamic equations}
Let us consider a droplet consisting of deconfined matter at about critical temperature. As model calculation, show, under such conditions the gluons acquire a large mass and are strongly suppressed by the Boltzmann factor \cite{Pes96, Lev98}. We take, therefore, only quarks and anti-quarks, in consideration. The quark-antiquark plasma inside the droplet is described by the hydrodynamic equations as a perfect fluid which is characterized by its rest frame energy density $\epsilon$, isotropic pressure $P$ and spherically-symmetric velocity field $\vec{v}(r)$ with $r$ being the distance to the center of the droplet.  For the equation of state, we use, the MIT bag model i.e. an ideal gas of massless quarks and antiquarks confined in the spherical cavity of radius R(t). The energy momentum tensor of the fluid is represented as
\begin{equation}
T^{\mu\nu} = T^{\mu\nu}_{\rm plasma} + T^{\mu\nu}_{\rm vacuum} =(\epsilon+P)u^\mu u^\nu - Pg^{\mu\nu} + Bg^{\mu\nu},
\end{equation}
where $g^{\mu\nu}$ is the metric tensor $g^{\mu\nu} = \text{diag}[1,-1,-1,-1]$ and $u^\mu = \gamma (1,\vec{v}(r))$ is the collective 4-velocity. The last term in this expression corresponds to the energy-momentum tensor of the "false" vacuum inside the droplet, characterized by the energy density $B$ (bag constant). Let us first consider an idealized case where the droplet does not emit particles from the surface. Then the dynamics of the fluid follows the energy-momentum conservation equation for $\nu=0$:
\begin{equation}
\partial_\mu T^{\mu 0} = \partial_0 T^{00} + \partial_i T^{i0} = 0, \qquad i=1,2,3. \label{EC}
\end{equation}
In the considered case the entropy is also conserved i.e. expressed by the equation:
\begin{equation}
\partial_\mu (su^\mu) = \partial_0 (su^{0}) + \partial_i (su^{i}) = 0, \label{SC}
\end{equation}
where $s$ is the entropy density. Furthermore, we can add the continuity equation expressing the conservation of quantum numbers such as baryon charge and strangeness:
\begin{equation}
\partial_\mu (n_j u^\mu) = \partial_0 (n_j u^{0}) + \partial_i (n_j u^{i}) = 0, \qquad j=B,S,\ldots \label{NC}
\end{equation}
where $n_j$ is the corresponding number density. From these equations, we will now derive the global dynamics of the droplet. Following ref. \cite{Bir83,Bar90} this can be done by averaging hydrodynamic equations over the spatial coordinates. All our calculations are done in the center of mass coordinate system and under the assumption that the temperature and all thermodynamical functions are homogenous within the droplet. For the velocity field of the fluid, we choose a Hubble-like profile of the form
\begin{equation}
\vec{v} = \frac{\vec{r}}{R} \cdot \dot{R}, \label{velfield}
\end{equation}
with $\dot{R}$ being the velocity of the surface of the droplet. A more general form of the flow profile as e.g. $\vec{v} = \left(\frac{\vec{r}}{R}\right)^{\alpha} \dot{R}$ (for positive $\alpha$), could be considered too (see e.g. ref. \cite{Pol98}), but for the sake of simplicity, we will adopt the linear form. With this choice we average Eqs. (\ref{EC}), (\ref{SC})  and (\ref{NC}) over the volume $V$ of the droplet. This yields for the l.h.s. of Eq. (\ref{EC})
\begin{equation}\label{EC2}
\begin{split}
\frac{1}{V} \int_{V}  \partial_\mu T^{\mu 0} \ dV &= \frac{1}{V} \int_{V} \frac{\partial}{\partial t} T^{00} \ dV + \frac{1}{V} \int_{V} \nabla \cdot \left[(\epsilon+P) \gamma^2 \vec{v}\right] \ dV  \\
&= \frac{1}{V} \frac{d}{dt} \left\{ V \left[ (\epsilon + P) \langle \gamma^2\rangle -  P +  B \right] \right\}.
\end{split}
\end{equation}
The second term in the r.h.s. of Eqs. (\ref{EC2}) vanishes because the volume integration can be replaced by an integral over the remote surface, where all thermodynamic quantities vanish. Analogously, for Eqs. (\ref{SC}) and (\ref{NC}), we have
\begin{equation} \label{SC2}
\begin{split}
\frac{1}{V} \int_{V}  \partial_\mu (su^\mu) \ dV &= \frac{1}{V} \int_{V}  \frac{\partial}{\partial t} (s \gamma)  \ dV + \frac{1}{V} \int_{V}  \nabla \cdot (s \gamma \vec{v}) \ dV \\
&= \frac{1}{V} \frac{d}{dt}  \left[ V s \langle \gamma \rangle \right],
\end{split}
\end{equation}
and
\begin{equation}\label{NC2}
\begin{split}
\frac{1}{V} \int_{V}  \partial_\mu (n_ju^\mu) \ dV &= \frac{1}{V} \int_{V}  \frac{\partial}{\partial t} (n_j \gamma)  \ dV + \frac{1}{V} \int_{V}  \nabla \cdot (n_j \gamma \vec{v}) \ dV\\
&= \frac{1}{V} \frac{d}{dt}  \left[ V n_j \langle \gamma \rangle \right].
\end{split}
\end{equation}
Eqs. (\ref{EC2}), (\ref{SC2}) and (\ref{NC2}) contain spatial averages of $\gamma$ and $\gamma^2$ (denoted as $\langle \gamma \rangle$ and $\langle \gamma^2 \rangle$). Using the parametrization in Eq. (\ref{velfield}), one can get the explicit expressions:
\begin{align}
\langle \gamma \rangle &= \frac{\int \gamma \ d^3r}{\int d^3 r}
= \frac{3}{2} \frac{1}{\dot{R}^3}\left(\arcsin(\dot{R})-\dot{R}\sqrt{1-\dot{R}^2}\right),\label{gamma}\\
\langle \gamma^2 \rangle &= \frac{\int \gamma^2 \ d^3r}{\int d^3 r} \label{gammaSQR}
 =\frac{3}{\dot{R}^3}\left(\text{arctanh}(\dot{R})-\dot{R}\right).
\end{align}
At $\dot{R}\rightarrow 0$, these expressions are reduced to
\begin{align}
\langle \gamma \rangle &\approx 1 + \frac{3}{10} \dot{R}^2 + \frac{9}{56} \dot{R}^4 + \cdots,  \qquad &\langle \gamma^2 \rangle \approx 1 + \frac{3}{5} \dot{R}^2 + \frac{3}{7} \dot{R}^4 + \cdots
\end{align}

Since the droplets are surrounded by a dilute hadron gas or even by the physical vacuum, there should be a surface energy associated with the change of the vacuum condensates inside the droplet. This means, we should add to Eq. (\ref{EC2}) a surface term, which is commonly disregarded in the hydrodynamical calculations. In our present calculations we use a standard parametrization of the surface energy with form $4 \pi R^2 \sigma$, where $\sigma$ is the surface tension coefficient. For the moving surface this expression has to be modified, because it should also contain the kinetic energy of the moving surface. The correct expression can be obtained by using an analogy with a relativistic particle of mass $4\pi R^2\sigma$ moving with velocity $\dot{R}$, i.e.
\begin{equation}
E_{\rm surface}(R,\dot{R}) = \frac{4\pi R^2 \sigma}{\sqrt{1-\dot{R}^2}}. \label{ensSurf}
\end{equation}  
Finally we obtain the following expressions for the total energy, entropy and any conserved quantum number in the droplet:
\begin{align}
& E(R,\dot{R}) = \frac{4\pi}{3} R^3 \left[ (\epsilon + P) \langle \gamma^2 \rangle- P + B \right] + \frac{4\pi R^2 \sigma}{\sqrt{1-\dot{R}^2}}, \\
& S(R,\dot{R}) = \frac{4\pi}{3} R^3 s \langle \gamma \rangle, \\
& N_{\rm j}(R,\dot{R}) = \frac{4\pi}{3} R^3 n_j \langle \gamma \rangle, \qquad j=B,S,\ldots
\end{align}
Additionally, we need explicit expressions for internal energy density, pressure and entropy density for the plasma inside the droplet. For massless quarks and antiquarks, these thermodynamic quantities as functions of temperature $T$ and chemical potential $\mu$ are expressed as 
\begin{align}
\epsilon(T,\mu)&= \frac{7 \pi^2}{120} \nu_q T^4 + \frac{\nu_q}{4} T^2\mu^2 + \frac{\nu_q}{8\pi^2} \mu^4 , \qquad P(T,\mu)=\frac{1}{3}\epsilon(T,\mu), \label{ensdens} \\ 
 s(T,\mu)&=\frac{7\pi^2}{90} \nu_q T^3 + \frac{\nu_q}{2}T\mu^2,\label{entdens}\\ 
 n_{\rm B}(T,\mu) &= \frac{\nu_q}{18} T^2\mu + \frac{\nu_q}{18}\frac{\mu^3}{\pi^2} \label{numdens},
\end{align}
where $\nu_{q}$ is the degeneracy factor of quarks (antiquarks). Since we include: up, down and strange quarks, $\nu_q = 18$ in our calculations. For strange quarks Eqs. (\ref{ensdens}), (\ref{entdens}) and (\ref{numdens}) should be generalized to include finite rest mass $m_s \approx 150$ MeV. But for the sake of simplicity, in our calculations, we will disregard the strange quark mass. Please note, that in the thermodynamic expressions below, the variables $T$ and $\mu$ are often suppressed.

\subsection{Choice of model parameters}
To perform further calculations, we should specify the model parameters and initial conditions for the droplet evolution. As already stated above, for calculation of the droplet bulk energy we use the MIT bag model. One should bear in mind, however, that in an equilibrated system this model predicts a first-order phase transition between the quark-gluon plasma and a hadron gas at all $T$ and $\mu$, see e.g. ref. \cite{Sat09}. On the other hand, the QCD based lattice calculations show a crossover type of the deconfinement  phase transition at small $\mu$. It occurs in the temperature interval (160-190) MeV, depending on the numerical scheme and quantity considered, for details see refs.\cite{Kar06,Fod09}. In a non-equilibrium situation, when the transformation from quark-gluon to hadronic degrees of freedom is not fast enough, a formation of quark droplets is quite likely to happen. As shown in refs. \cite{Mis99b,Mis08}, the QGP phase produced in relativistic heavy-ion collisions may split into droplets due to a strong collective expansion. Below, we assume that deconfinemed matter inside these droplets may exist in a metastable state at temperatures as low as 150 MeV and $\mu \approx 0$. To describe this state we apply the bag model with bag constant $B = 200 \ \rm \frac{MeV}{fm^3}$, which is slightly lower than the standard value around $235 \ \rm \frac{MeV}{fm^3}$. With this choice, the balance between the plasma pressure (\ref{ensdens}) and the vacuum pressure ($P_{\rm vac}= -B$) is achieved at a temperatures $T_0 \approx 150$ MeV.

For small droplets, the finite-size corrections to the droplet energy should be taken into account in addition to the bulk term proportional to $R^3$. As shown in ref. \cite{Mar91}, within the MIT bag model with massless quarks, the standard surface term proportional to $R^2$ vanishes, and the curvature term proportional to $R$ has a negative sign. The implications of this term for finite-size quark-gluon droplets in the mixed phase of the deconefinement phase transition, were studied in ref. \cite{Bri98}. The author's conclusion is that this correction is rather small (around 30 \%), even for droplet radii about 1 fm, and negligible for radii $R \gtrsim 2 \ \rm fm$.

The MIT bag model assumption of an idealized sharp boundary of the quark bag might not be realistic enough to correctly describe the surface properties. Other approaches, which are based on effective field-theorectical models, usually predict a non-vashing surface energy. This energy is associated with the gradients of the mean fields (condensates) in the transition region between inside and outside of the droplet. For instance, the calculations within the linear sigma model with quarks, ref. \cite{Fra11}, predict the surface tension coefficient in the range $(5-15) \ \rm \frac{MeV}{fm^2}$. Other estimates using lattice QCD results \cite{Bug10,Dum11} give even larger values, of order of 100 $\rm \frac{MeV}{fm^2}$. Taking into account these uncertainties, in our exploratory study, we choose the surface energy in the form of Eq. (\ref{ensSurf}) with temperature-independent coefficient $\sigma = 50 \ \rm \frac{MeV}{fm^2}$.

For the initial droplet radius we take the values of about 2 fm, which is motivated by estimates of ref. \cite{Mis09}. There the droplet size was related to the expansion rate of the quark-gluon plasma just before the break-up. It is clear, that this primordial plasma expansion should lead to the residual droplet expansion, hence in our investigation we also consider droplets with initial radial velocities, i.e. $\dot{R}_0 \approx 0.2 c$

\section{Droplet dynamics without hadron emission}
\subsection{Energy functional}
\begin{figure}[ht!]
\centering
\includegraphics[width=\linewidth]{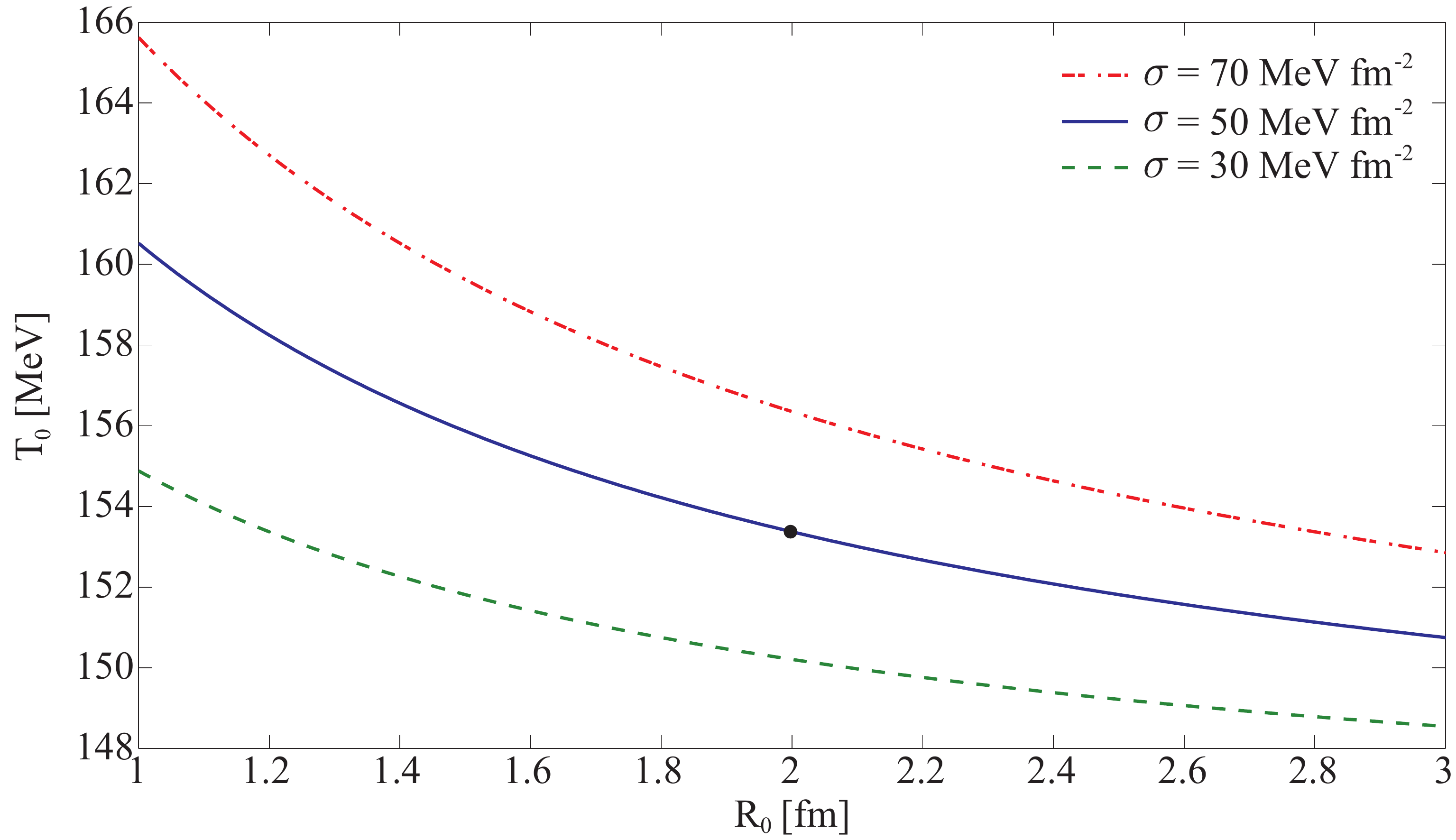}
\caption{\small The equilibrium radius of the droplet as a function of temperature calculations for $\sigma=30, 50$ and $70 \ \rm \frac{MeV}{fm^2}$ and $B = 200 \ \rm \frac{MeV}{fm^3}$. In our calculations, we primarily choose (indicated by the dot), $\sigma=50 \ \rm \frac{MeV}{fm^2}$, $B = 200 \ \rm \frac{MeV}{fm^3}$ and $R_0= 2$ fm which corresponds to a temperature of $153.4$ MeV.}\label{plot:equilibrium}
\end{figure}

When hadron emission from the droplet is disregarded, its energy, entropy and net baryon number are conserved i.e. 
\begin{align}
E(R,\dot{R}) &= E_0, \label{a}\\
S(R,\dot{R}) &= S_0, \label{b}\\ 
N_{\rm B}(R,\dot{R}) &=  N_{\rm B, 0}.
\end{align}
In the simplest case of zero chemical potential, we are left with expressions of the energy and entropy densities as functions of temperature only. Then Eq. (\ref{b}) can be used to eliminate the temperature from Eq. (\ref{a}) leading to an expression containing $R$ and $\dot{R}$ in a very nonlinear way. The resulting energy functional is 
\begin{align}
E(R,\dot{R}) &=  \frac{\kappa}{R} \frac{4\langle \gamma^2\rangle-1 }{3\langle \gamma \rangle^{4/3}} +\frac{4}{3} \pi R^3 B + \frac{4\pi R^2\sigma}{\sqrt{1-\dot{R}^2}},\label{EnFunc}
\end{align}
where $\langle \gamma \rangle$ and $\langle \gamma^2 \rangle$ are given by Eqs. (\ref{gamma}) and (\ref{gammaSQR}), and $\kappa=\frac{3}{4\pi} \left[\frac{135 S_0^4}{14 \nu_q}\right]^{1/3}$. Inserting this expression in Eq. (\ref{a}) leads to a nonlinear differential equation, whose direct solution is difficult to obtain. To solve it, we need further simplifications, which are introduced below. First, let us analyze the shape of the energy functional (\ref{EnFunc}). The equilibrium radius of the droplet, $R_0$, corresponds to the minimum energy state at $\dot{R}$, i.e.
\begin{equation}
\left. \frac{\partial E}{\partial R}\right|_{R=R_0, \dot{R}=0} = - \frac{\kappa}{R_0^2} +4 \pi R_0^2 B + 8 \pi R_0 \sigma = 0. \label{eq1}
\end{equation}
This condition is equivalent to the pressure balance at the surface
\begin{equation}
P_0(T_0) = B + \frac{2\sigma}{R_0}, \label{constrain}
\end{equation} 
where $P_0$ is the pressure of the quark gas inside the droplet, $B$ is the vacuum pressure, and $\frac{2\sigma}{R_0}$ is the Laplace pressure. The numerical solution of Eq. (\ref{constrain}) is presented in Fig. \ref{plot:equilibrium} for three different values of $\sigma$.  For the default values of $B$ and $\sigma$, an equilibrium radius of 2 fm requires the initial temperature of 153.4 MeV. Then the minimal energy of the droplet is around 34.4 GeV. Obviously, to have higher initial temperature with the same radius, one should increase $\sigma$ or $B$.

\begin{figure}[ht]
\centering
\includegraphics[width=\linewidth]{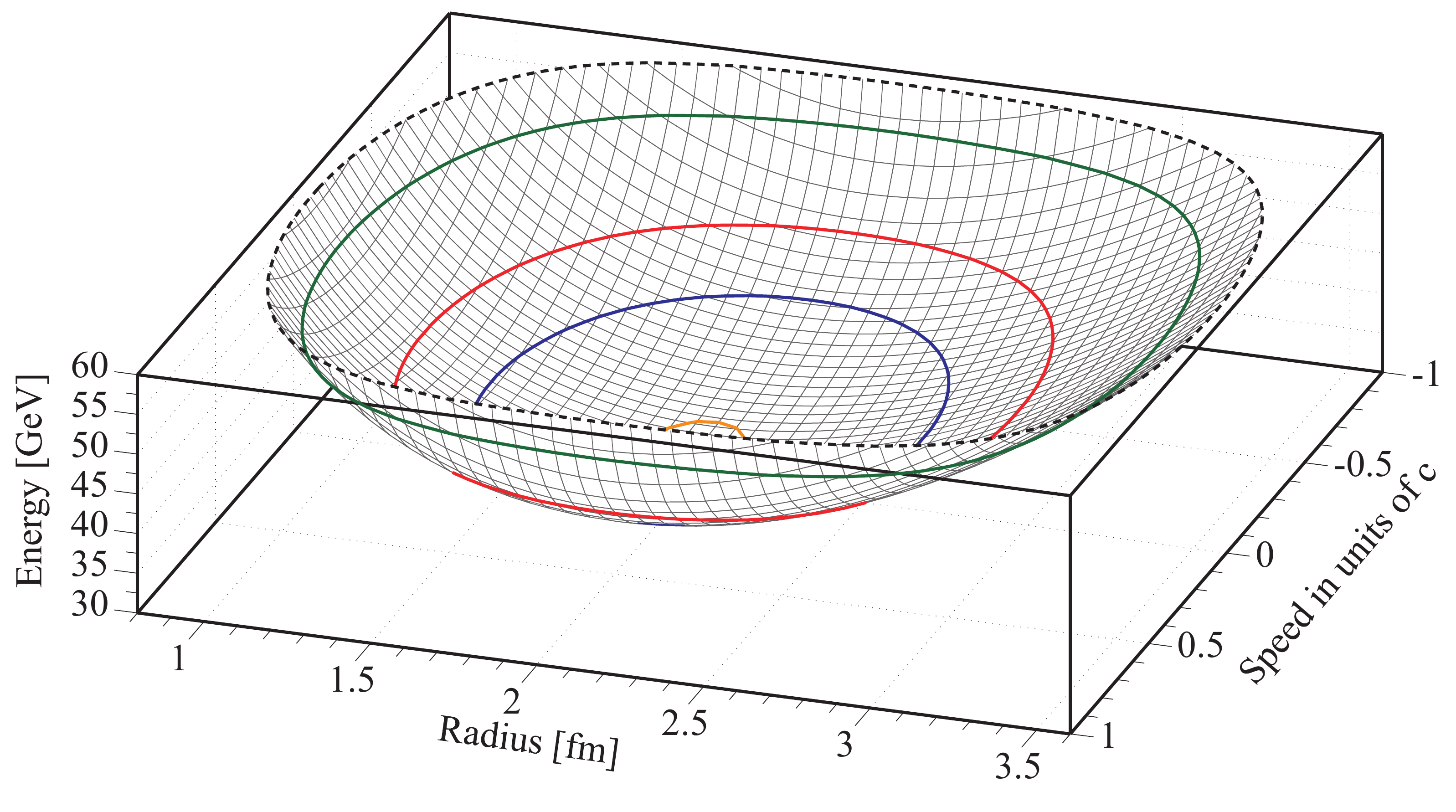}
\caption{\small The droplet energy as a function of its radius and surface speed. Calculations are done for initial temperature: $T_0=154.3 \ \rm MeV$ and model parameters $\sigma=50 \ \rm \frac{MeV}{fm^2}$ and $B=200 \ \rm \frac{MeV}{fm^3}$. The minimum corresponds to about 34.4 GeV at radius 2 fm. The colored contours corresponds to the iso-energetic contours at four different energies: 34.6, 39.6, 44.6 and 54.6 GeV shown in the next figure.}\label{plot:fullEllipse}
\end{figure}
\begin{figure}[ht]
\centering
\includegraphics[height=8cm]{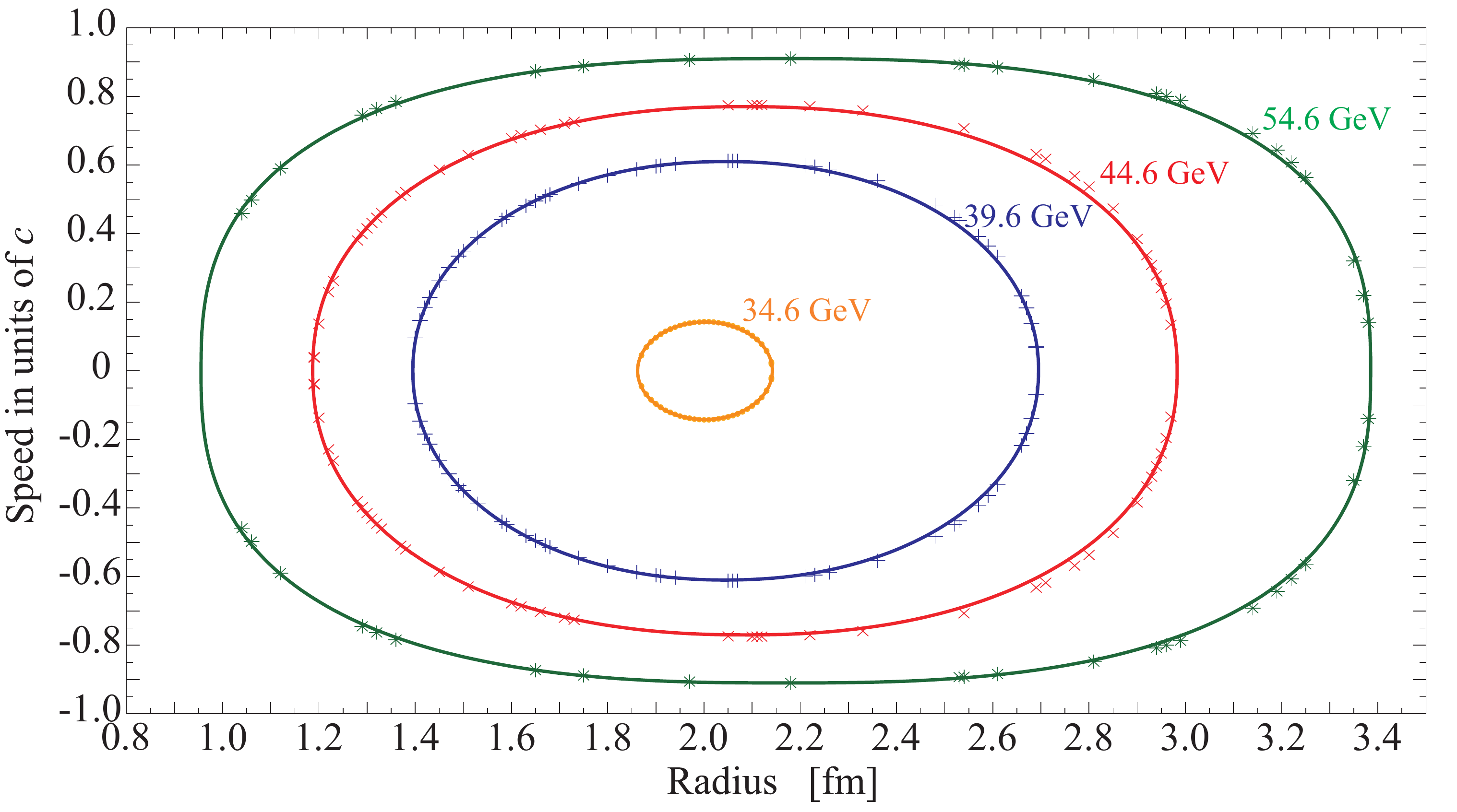}
\caption{\small Iso-energetic contours in ($R$, $\dot{R}$) plane shown for four different energies (34.6, 39.6, 44.6 and 54.6 GeV) together with fits to super ellipses (Eq. \ref{superfit}). The parameters for the fits can be found in Table \ref{abnparameters}.}\label{plot:SuperEl}
\end{figure}

For further calculations, it is useful to make a graphical representation of the energy functional in phase space defined by variables $R$ and $\dot{R}$. The corresponding 3-dimensional plot is given in Fig. \ref{plot:fullEllipse}. As one can see, the energy surface is symmetric in the velocity coordinate $\dot{R}$, while it is rather asymmetric in the $R$ coordinate. The horizontal cuts of this surface at $E_0=$ 34.6, 39.6, 44.6 and 54.6 GeV are shown in Fig. \ref{plot:SuperEl}. For lower energies, they look rather symmetric with respect to the point ($R_0$, 0) both in radius and velocity coordinates, while at higher energies, one can notice deviations at small radii. 

Rather surprisingly it turns out that these contours fit quite well with the shape of superellipses (see ref. \cite{Gar77}), defined by the equation
\begin{equation}
\left| \frac{R-R_c}{a} \right|^n + \left| \frac{\dot{R}}{b} \right|^n = 1, \label{superfit}
\end{equation}
where $R_c$, $a$, $b$ and $n$ are fitting parameters. These fits are shown together with the data points in Fig. \ref{plot:SuperEl}, and the corresponding parameters are listed in Table \ref{abnparameters}. 

\begin{figure}[ht]
\centering
\includegraphics[width=\linewidth]{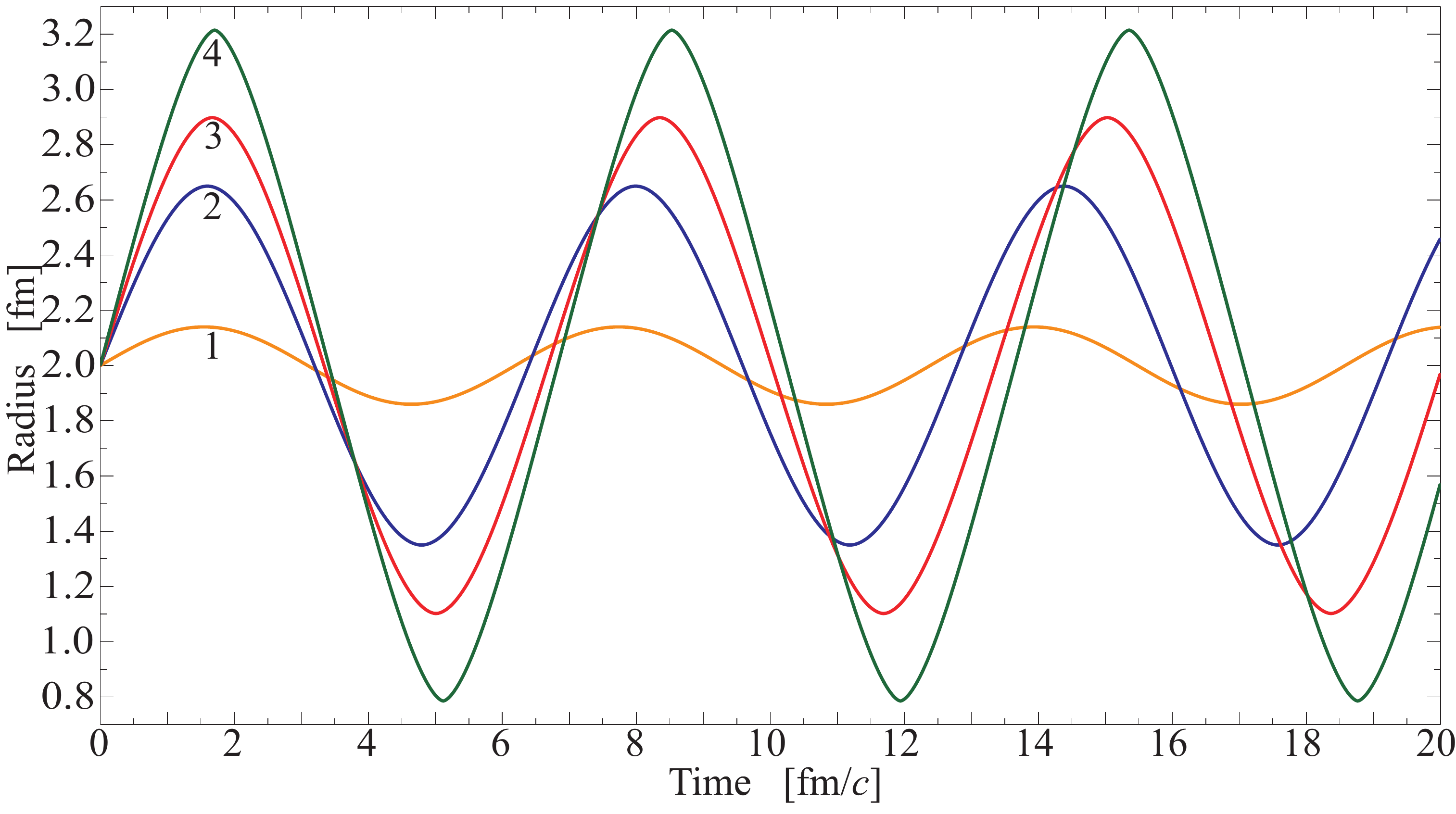}
\caption{\small Numerical solutions for the radius of the droplet as a function of time obtained after fitting the iso-energetic contours to super ellipses. The coloring matches the previous plot. The curves labeled with numbers 1,2,3 and 4 correspond to energies 32.5, 37.5, 42.5 and 52.5 GeV respectively.}\label{plot:radiusVStime}
\end{figure}

With parametrization of Eq. (\ref{superfit}), we can easily find a numerical solution for the time-dependent radius of the droplet using standard methods like the Runge-Kutta scheme. In Fig. \ref{plot:radiusVStime}, the corresponding solutions are presented for the same total energies as in Fig. \ref{plot:SuperEl}. 
The radial oscillations of droplets are explained as follows. Since we start with the same radius $R_0=$ 2 fm and temperature $T_0=153.4 \ \rm MeV$, the different total energies correspond to different initial velocities; around 0.14 for the droplet with smallest energy and 0.91 for the droplet with the biggest energy (see Eq. \ref{EnFunc}). A droplet with an outward velocity will expand isentropically until the temperature of the thermal gas drops to a level where its pressure becomes smaller than the sum of the bag pressure and the Laplace pressure (see Eq. (\ref{constrain})). At this point, the droplet begins to contract until the plasma inside the droplet is heated up sufficiently to increase the thermal pressure. Generally, these solutions describe anharmonic oscillations of the droplet around the minimum energy state. The anharmonicity effects are getting more and more significant with increasing energy. The deviations from the harmonic oscillations are clearly seen in the shape of the curves in Fig. \ref{plot:radiusVStime}. The periods of the ocillilations are listed in Table \ref{abnparameters}.

\begin{table}[htdp]
\begin{center}
\caption{\small Fitting parameters: $a$,$b$ and $n$ for the super ellipse equation together with the coordinate $R_c$ at the center and the period of oscillation.} \label{abnparameters}
\begin{tabular}{|ccccccc|}\hline
$E$, [GeV] & $a$, [fm] & $b$, [$c$] & $n$ & $R_c$, [fm] & $\dot{R}_0$, [$c$] & $\tau$, [fm/$c$]  \\ \hline  \hline
34.6 & 0.14 & 0.14 & 2.00 & 2.00 & 0.14 & 6.19 \\ \hline 
39.6 & 0.65 & 0.61 & 2.10 & 2.04 & 0.61 & 6.39 \\ \hline 
44.6 & 0.90 & 0.77 & 2.22 & 2.08 & 0.77 & 6.68 \\ \hline 
54.6 & 1.21 & 0.91 & 2.65 & 2.17 & 0.91 & 6.82 \\ \hline 
\end{tabular}
\end{center}
\label{default}
\end{table}

\subsection{Small amplitude oscillations}
It is instructive to compare numerical results presented above with an analytical solution, which can be obtained in the case of small-amplitude oscillations. Expanding the energy functional of Eq. (\ref{EnFunc}) around the equilibrium point $R=R_0$, $\dot{R}=0$, we get 
\begin{align}
E(R,\dot{R}) &= \frac{\kappa}{R_0} \left[3- 3x + x^2 \right] \left[1+ \frac{2}{5} \ (R_0\dot{x})^2 + \cdots\right]\\
& + \frac{4}{3} \pi R_0^3 B \left[ 1 + 3 x + 3 x^2 + \cdots \right] + 4\pi R_0^2\sigma \left[ 1 + 2x+x^2\right] \left[1+\frac{1}{2} (R_0\dot{x})^2+\cdots\right],\nonumber 
\end{align}
where $x=\left(\frac{R}{R_0}-1\right)$. Retaining only quadratic terms in $x$ and $\dot{x}$, we get
\begin{equation}
E(R,\dot{R}) = E(R_0,0) + C_1x^2 + C_2R_0^2\dot{x}^2,
\end{equation}
where $C_1= \frac{\kappa}{R_0}+ 4\pi R_0^3  B + 4\pi R^2_0 \sigma$ and $C_2=\frac{3}{10}\frac{\kappa}{R_0} + 2\pi R_0^2 \sigma$. 
In this approximation the energy functional takes the form of a harmonic oscillator with angular frequency
\begin{equation}
\omega = \sqrt{\frac{C_1}{C_2}} \frac{1}{R_0} = \sqrt{5\frac{1-\frac{1}{2}\xi}{1+\frac{5}{4}\xi}} \frac{1}{R_0}, \qquad \xi= \frac{E_{\rm surface}}{E_{\rm plasma}} = \frac{4\pi}{\kappa} R_0^3 \sigma,
\end{equation}
where $E_{\rm surface} = 4 \pi R_0^2  \sigma$ and $E_{\rm plasma} = \frac{\kappa}{R_0}$. Now we can estimate the oscillation period for the considered droplet ($E_{\rm surface}= 2.5 \ \rm GeV$, $E_{\rm plasma}= 25.1 \ \rm GeV$ ) as
\begin{equation}
\tau = \frac{2\pi}{\omega} = 6.11 \ \rm \frac{fm}{c}. \label{OSCtime}
\end{equation}
This fits well with our numerical solution for the lowest energy (Fig. \ref{plot:SuperEl}), which shows the period of 6.19 fm/$c$. At higher initial energies anharmonic effects become apparent and the oscillation period increases.

\section{Statistical description of hadron emission}
In this section, we consider the hadron emission from the droplet's surface. The idea is that the quarks close to the surface can combine into colorless hadrons, which may leave the droplet carrying away the energy $\omega_h$. We will describe this process following Weisskopf's statistical model in \cite{Wei37}. A similar approach has been used earlier in ref. \cite{Bar90}. The ability to emit hadrons is determined by the thermal excitation energy $E^\star(T,R,\dot{R})= E(T,R,\dot{R})-E(0,R,\dot{R})$. The droplet is assumed to be spherical at all times and in thermal equilibrium before and after each emission. According to Weisskopf's model, the double-differential emission rate is given by the ratio of the statistical weights before and after the emission, $ \frac{\Omega_{\rm after}}{\Omega_{\rm before}}$:
\begin{figure}[ht]
\centering\includegraphics[width=\linewidth]{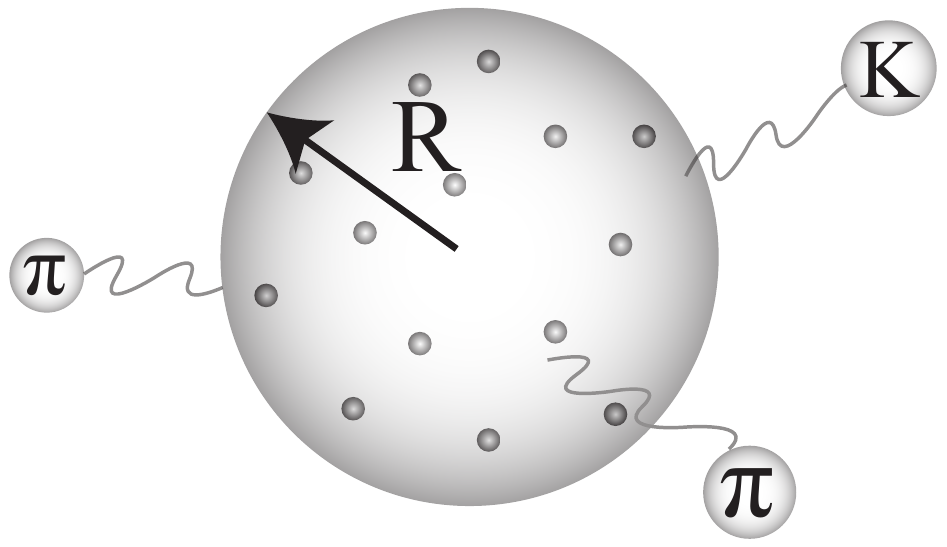}
\caption{\small Hadron emission from the surface of a quark droplet of radius $R$.}\label{pics:HE}
\end{figure}
\begin{equation}
\begin{split}
\frac{d^2N_h}{dp\ dt} &= \frac{1}{2} \frac{\nu_h}{2 \pi^2} p^2 \mathcal{A} \exp[S(E^\star,N,R)-S(E^\star_0,N_0,R_0)] \\
&= \frac{1}{2} \frac{\nu_h}{4 \pi^2} p^2 \mathcal{A} \exp[\Delta S],
\end{split}
\end{equation}
where $p$ is momentum of emitted hadron, $\nu_h$ is its spin-isospin degeneracy factor, $\mathcal{A}$ is often estimated as the cross section of inverse reaction, but for our problem it is more appropriate to identify it with the total surface area of the droplet, $\mathcal{A} = \mathcal{A}_{\rm geom} = 4 \pi R^2$, $S(E^\star,R)$ is the entropy of the droplet at the thermal excitation energy $E^\star$ and volume $V=\frac{4}{3}\pi R^3$. Note that we have introduced a factor of 1/2, since the particle emission is only possible for polar angles $\theta < \frac{\pi}{2}$. Furthermore, in the case of a moving surface, the emission rate will be suppressed by the $\gamma$ factor \cite{Bar90}. The energy conservation in the emission process can be expressed as:
\begin{equation}
\Delta E_h = E -E_0 = - \omega_h, \qquad \omega_h = \sqrt{m_h^2 + p^2},
\end{equation}
where $m_h$ is the hadron mass. The change in the entropy due to emission of a hadron $h$ can be found from the 2nd law of thermodynamics:
\begin{equation}
\Delta S_h = \frac{1}{T} \left(\Delta E^\star_h + P \Delta V_h - \mu_h \Delta N_h \right),
\end{equation}
where $\Delta E_h= -\omega_h$, $P$ is the thermal pressure and $\Delta V_h$ is the change of the volume due to emission of one hadron, which will be neglected below, $\Delta N_h = -1$ since one particle is emitted and $\mu_h$ is the chemical potential of the droplet corresponding to the quantum numbers of the emitted hadron.  
Now, the double-differential emission rate can be written as
\begin{equation}
\frac{d^2N_h}{dp\  dt} = \frac{\nu_h}{4\pi^2} p^2 4\pi R^2 \exp \left[- \frac{\sqrt{m_h^2 + p^2}}{T} + \frac{\mu_h}{T} \right].
\end{equation}
The first two moments of this distribution give the total particle number and energy emission rates
\begin{align}
\frac{dN_h}{dt} &= \frac{\nu_h}{\pi} R^2 \int^\infty_0 p^2 e^{-\frac{\sqrt{m_h^2+p^2}-\mu_h}{T}} \ dp,\\ 
\frac{d E_{\rm loss}}{dt} &= \sum_h \frac{\nu_h}{\pi} R^2 \int^\infty_0 p^2 \sqrt{m_h^2 + p^2} e^{-\frac{\sqrt{m_\pi^2+p^2}-\mu_h}{T}} \ dp. \label{averageEloss}
\end{align}
In our calculations, we include emission of mesons ($\pi, K, \rho, \omega$), baryons $(N, \Delta)$ and hyperons ($Y=\Lambda, \Sigma$). The Fermi and Bose statistics effects are disregarded in these calculations. Obviously, the energy loss due to emitted particles leads to a change of the entropy and total energy of the droplet. Therefore, instead of Eqs. (\ref{a}) and (\ref{b}) we should now solve the following equations
\begin{align}
\frac{dE}{dt} &= \frac{d}{dt} \left[ V \{ (\epsilon+P)\langle \gamma^2 \rangle - P + B\} + \frac{4\pi R^2 \sigma}{\sqrt{1-\dot{R}^2}}\right] = - \frac{dE_{\rm loss}}{dt},\label{Eloss}\\ 
\frac{dS}{dt} &= \frac{d}{dt}\left[Vs \langle \gamma \rangle \right] = - \frac{1}{T} \frac{dE_{\rm loss}}{dt}. \label{Sloss}
\end{align}
These equations together with Eq. (\ref{averageEloss}) should be solved for functions $R(t)$ and $T(t)$.
\section{Numerical simulations of droplet dynamics}
To solve Eq. (\ref{averageEloss}), (\ref{Eloss}) and (\ref{Sloss}), we have created a numerical scheme calculating at each time step ($n\cdot \Delta t$) the change of the variables $R, \dot{R}$ and $T$ based on their previous values. We are using the Runge-Kutta method to determine $R_{n+1}$ and $\dot{R}_{n+1}$ from $R_n$, $\dot{R}_n$ and $\ddot{R}_{n}$. The acceleration $\ddot{R}_{n}$ can be calculated by knowing $S_{n}$, $R_n$ and $\dot{R}_n$. Now $T_{n+1}$ can be calculated from $S_n$, $R_{n+1}$ and $\dot{R}_{n+1}$. Finally, from calculating the energy loss from $T_{n+1}$ and $R_{n+1}$, we get $S_{n+1}$. Now the loop can be redone at the next time step: $(n+1)\cdot \Delta t$.
We have performed numerical simulations for four initial conditions, all with $R_0$= 2 fm : 1) $\dot{R}_0$=0, $T_0$ = 153.4 MeV (close-to-equilibrium case), 2) $\dot{R}_0$=0.2, $T_0$ = 153.4 MeV, 3) $\dot{R}_0$=0, $T_0$ = 140 MeV and 4) $\dot{R}_0$=0, $T_0$ = 175 MeV.

\subsection{Evolution of droplets with hadron emission}
\begin{figure}[ht!]
\centering
\includegraphics[width=\linewidth]{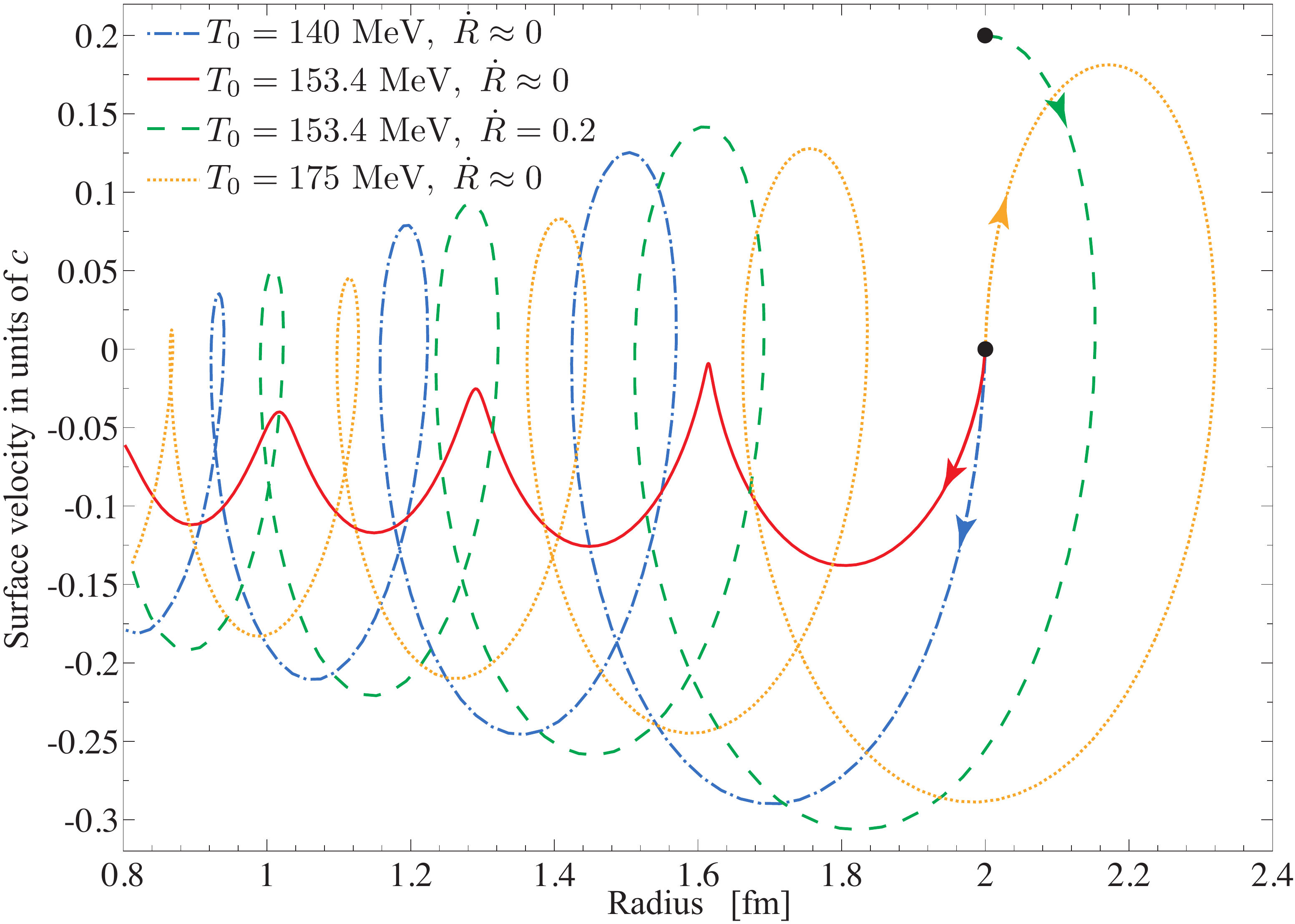}
\caption{\small Trajectories of the quark droplets in the $R-\dot{R}$ plane are shown for four different initial condition (indicated in the figure). A cut off at $R=0.8$ fm is applied. The dots and the arrows on the plot indicates the initial points and the direction of the droplet evolution.}\label{plot:velVSrad}
\end{figure}

The behavior of the droplets with hadron emission differ significantly from that in the emission-free case. In Fig. \ref{plot:velVSrad}, the surface velocity is plotted as a function of radius of the droplet, analogous to Fig. \ref{plot:radiusVStime} for the emission-free case. The dynamics shows a rather peculiar oscillating behavior superposed with the shrinkage of the droplet. For the close-to-equilibrium case without initial speed, the droplet does not acquire any outward speed at all. In other words, the general trend is a shrinkage of the droplet due to the hadron emission. Nevertheless, in the case of high initial temperature or non-zero initial speed, the droplet is initially expanding until $\dot{R}=0$ at radii of about 2.32 fm and 2.16 fm, respectively, and then shinks again.

\begin{figure}[ht]
\centering
\includegraphics[width=\linewidth]{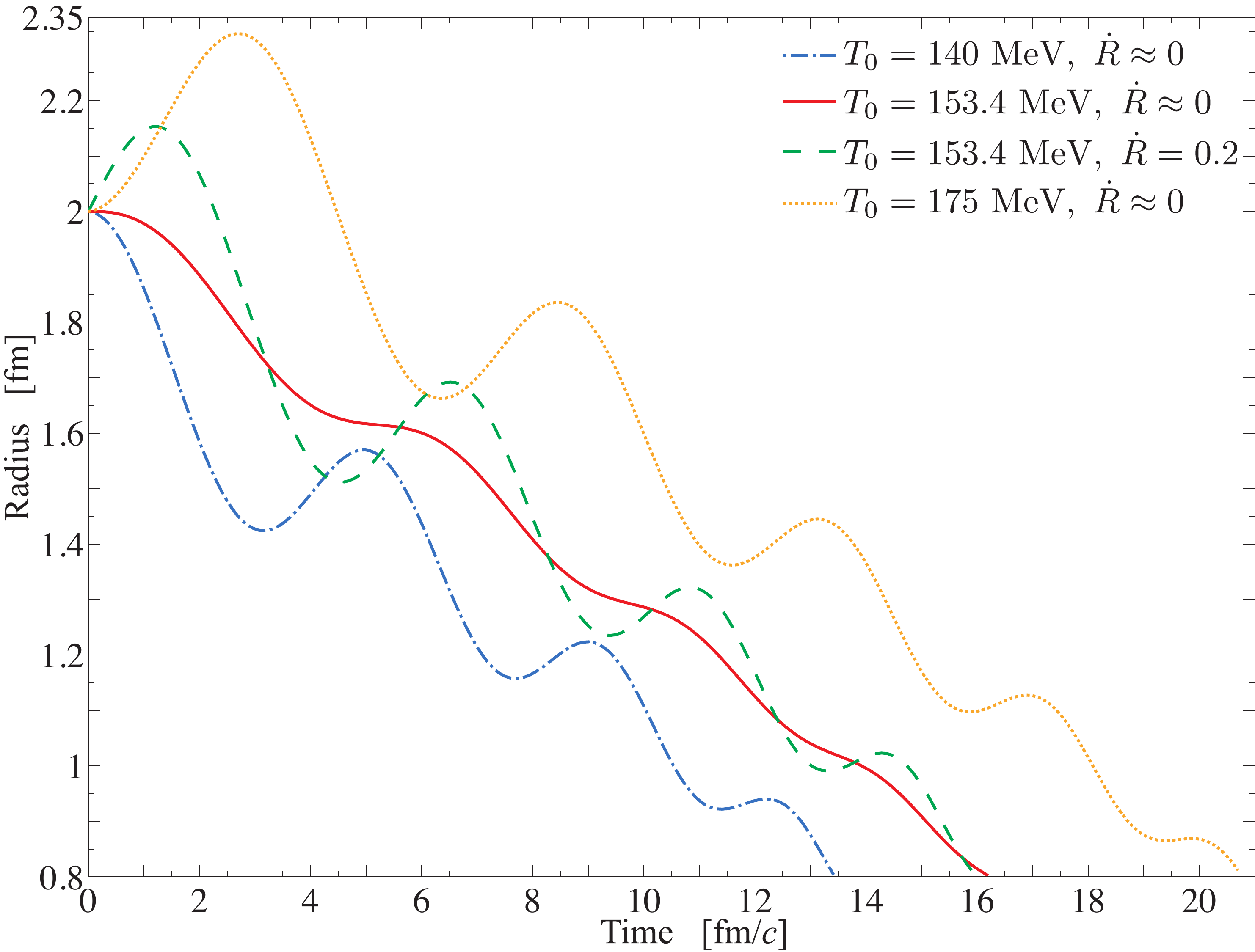}
\caption{\small The droplet radius as a function of time for four different initial conditions (indicated in the figure). A cutoff of $R$=0.8 fm is applied.}\label{plot:RadiusVStime}
\end{figure}

Fig. \ref{plot:RadiusVStime}, shows the radius of the droplet as a function of time. The solid line, corresponding to the close-to-equilibrium initial state shows, a gradual decrease of the radius due to the hadron emission. When the initial state is out of equilibrium (3 other curves), the droplet's radius exhibits damped oscillations, reminiscent of the droplet oscillations in the emission-free case (see Fig. \ref{plot:radiusVStime}). In all cases, after averaging over oscillations, the radius drops almost linearly with a rate of about 0.075 $c$.
We stop our calculations at radii of about 0.8 fm, close to the charge radius of the proton, because at smaller values our macroscopic considerations cannot be justified. As one can see, from Fig. \ref{plot:RadiusVStime}, a droplet with initial radius of 2 fm shrinks to this size in about 16 fm/$c$. This is of the order of the hadronization times extracted from the HBT measurements, which are about 10 fm/$c$ at LHC and around 7 fm/$c$ at RHIC \cite{Aam11}. However, one should bear in mind that the life time of the droplet is very sensitive to its initial radius. For instance, a droplet with radius of 1.5 fm will hadronize within the time interval of about 9 fm/$c$.

\begin{figure}[ht]
\centering
\includegraphics[width=\linewidth]{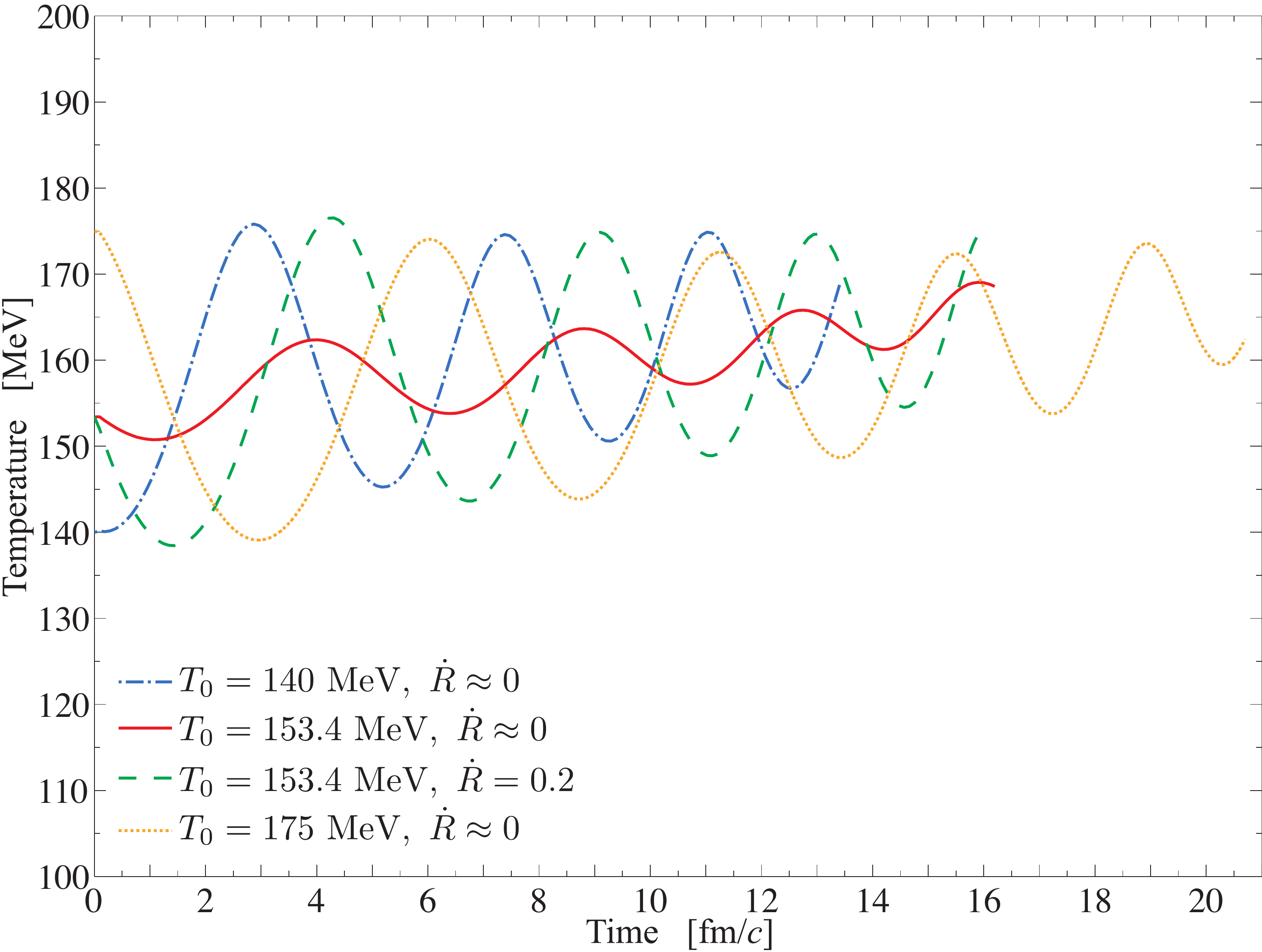}
\caption{\small The droplet temperature as a function of time for four different initial conditions (indicated in the figure). A cutoff at $R$=0.8 fm is applied.}\label{plot:TempVStime}
\end{figure}

In Fig. \ref{plot:TempVStime}, the temperature of the quark gas inside the droplet is plotted as a function of time for the same four initial conditions as discussed above. One can see, that for the high-temperature case and the non-zero initial speed case the temperature of the quark gas decreases and reaches around 140 MeV at the turning point for both cases. At that point, the thermal pressure is too low to balance the vacuum pressure and the Laplace pressure, and the droplet therefore starts to contract until the temperature reaches a new maximum around 175 MeV for both cases.

We can make two conclusions from Fig. \ref{plot:TempVStime}: First, the temperature of the droplet slowly rises during its life time, and second, oscillations of the temperature get smaller with time. This shows that due to the damping, the temperature needs to increase slightly to balance the surface tension of the smaller droplet i.e. the droplets are evolving towards an equilibrium state, which however, is constantly changing to higher temperatures according to the curve depicted in Fig. \ref{plot:equilibrium}. This trend is different from the one found in ref. \cite{Bar90}, where the surface tension was disregarded, so that the droplet temperature was approximately constant.

\subsection{Characteristics of emitted hadrons}
\begin{figure}[ht]
\centering
\includegraphics[width=\linewidth]{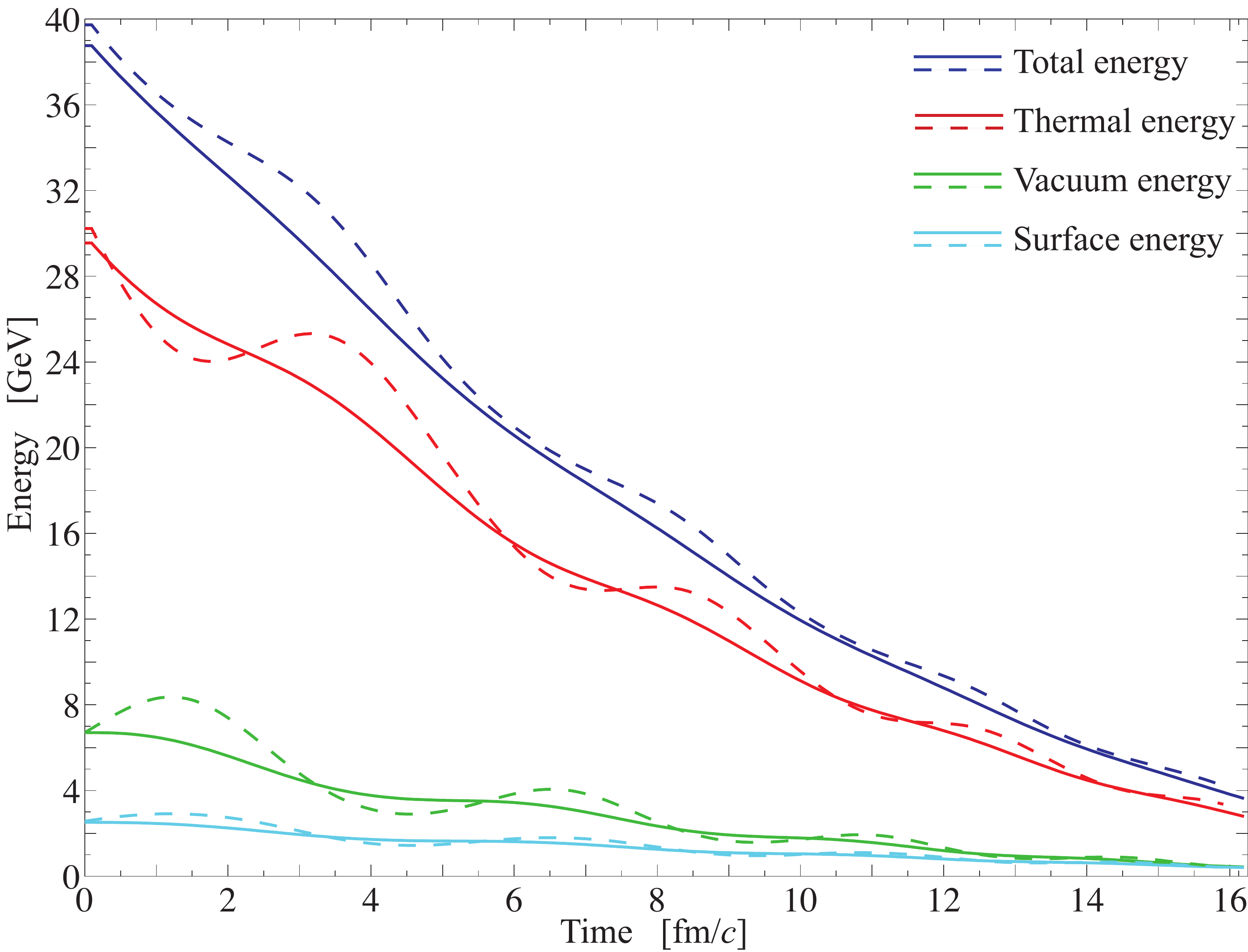}
\caption{\small Time evolution of different contributions to the total energy of the droplet, the thermal energy of the quark gas, the surface energy and the vacuum energy due to the bag constant. The solid and dashed lines represent initial conditions with $\dot{R}_0 = 0$ and $\dot{R}_0 = 0.2$ respectively.}\label{plot:Energyplot}
\end{figure}

Now, let us discuss characteristics of the emission process. We start with the energy budget of the droplet. Fig. \ref{plot:Energyplot} shows, the total energy of the droplet, the thermal energy of the quark gas, the surface energy and the vacuum energy as functions of time for two different initial conditions i.e $\dot{R}_0 = 0$ (solid lines)  and $\dot{R}_0 =0.2$ (dashed lines) at the same temperature $T_0 =$153.4 MeV. The two cases differs only by around 1 GeV, since the difference in initial speed  is rather small. At all times, the dashed line lies above the solid line for the total energy as we would expect. In the case of non-zero initial speed, we see a more violent dynamics in accordance with the earlier plots. In both cases, we have an initial energy of about 40 GeV which drops to around 4 GeV during the lifetime of the droplet. Most of the energy is accumulated in the thermal gas; about 30 GeV out of 40 GeV at initial stage. One can also see, that the oscillations of the total energy are coupled to the variations of the thermal energy. The oscillations of the vacuum energy are entirely due to the oscillations of the droplet volume, while the oscillations of the surface energy are caused by variations of the surface area and the speed.

\begin{figure}[ht]
\centering
\includegraphics[width=\linewidth]{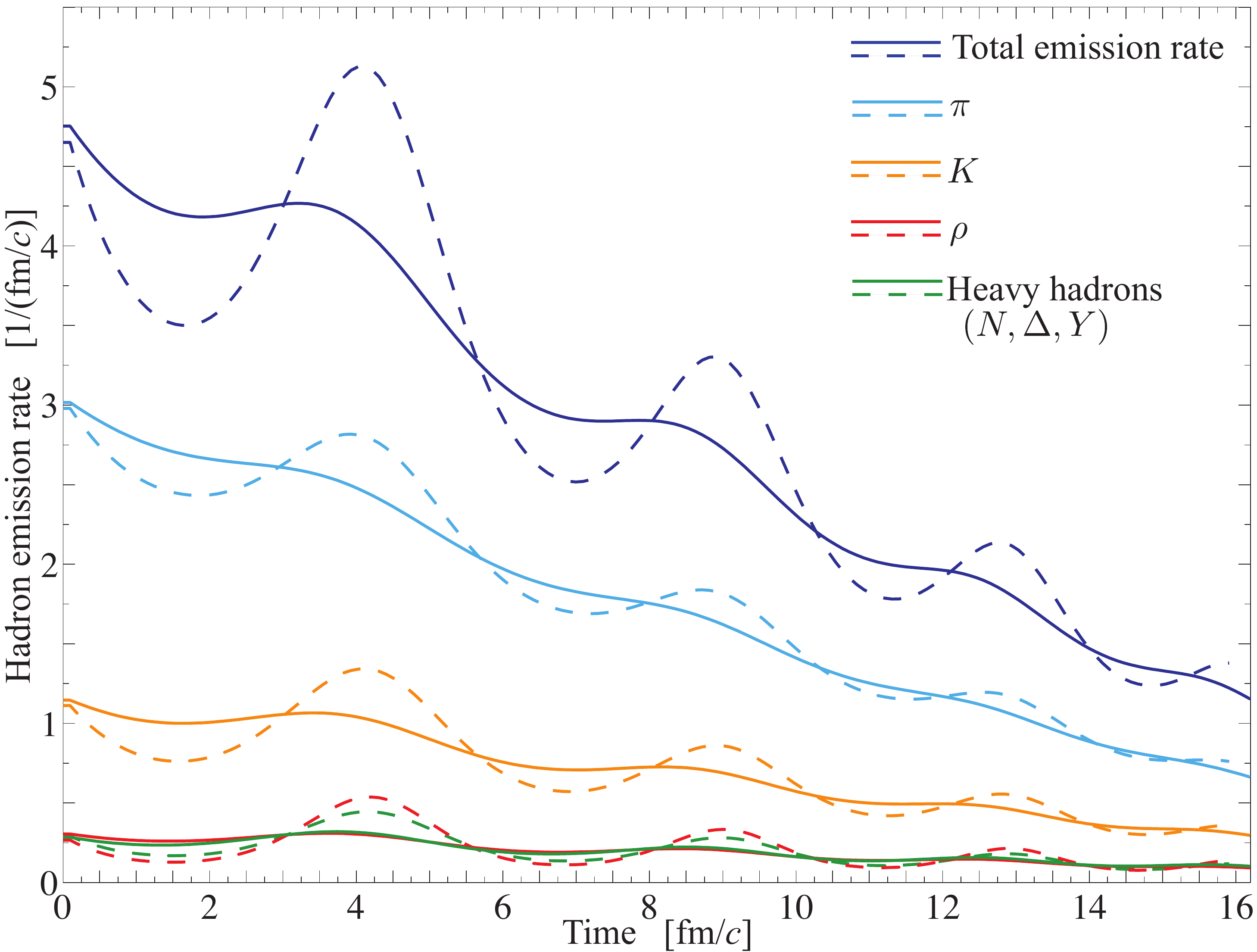}
\caption{\small Emission rates of different hadron species (indicated on the figure) and the total emission rate as function of time for two different initial conditions. The solid and dashed lines represent calculation with $\dot{R}_0 = 0$ and $\dot{R}_0 = 0.2$ respectively. }\label{plot:Hadrons2}
\end{figure}

\begin{figure}[ht]
\centering
\includegraphics[width=\linewidth]{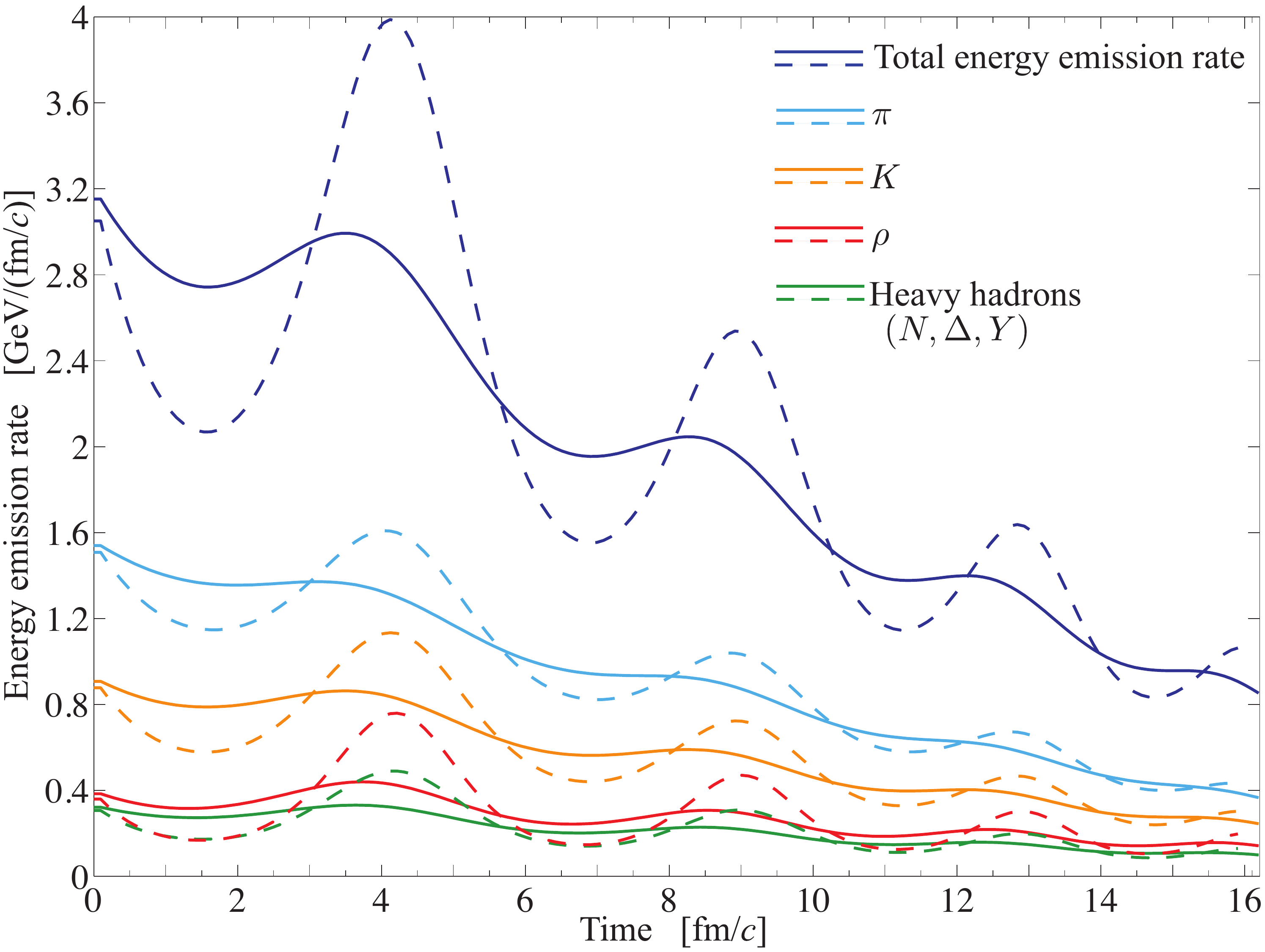}
\caption{\small Energy emission rates for different hadron species (indicated on the figure) and the total energy emission rate as function of time for two different initial conditions. The solid and dashed lines represent calculation with $\dot{R}_0 = 0$ and $\dot{R}_0 = 0.2$ respectively.}\label{plot:Energy2}
\end{figure}

\begin{figure}[ht]
\centering
\includegraphics[width=\linewidth]{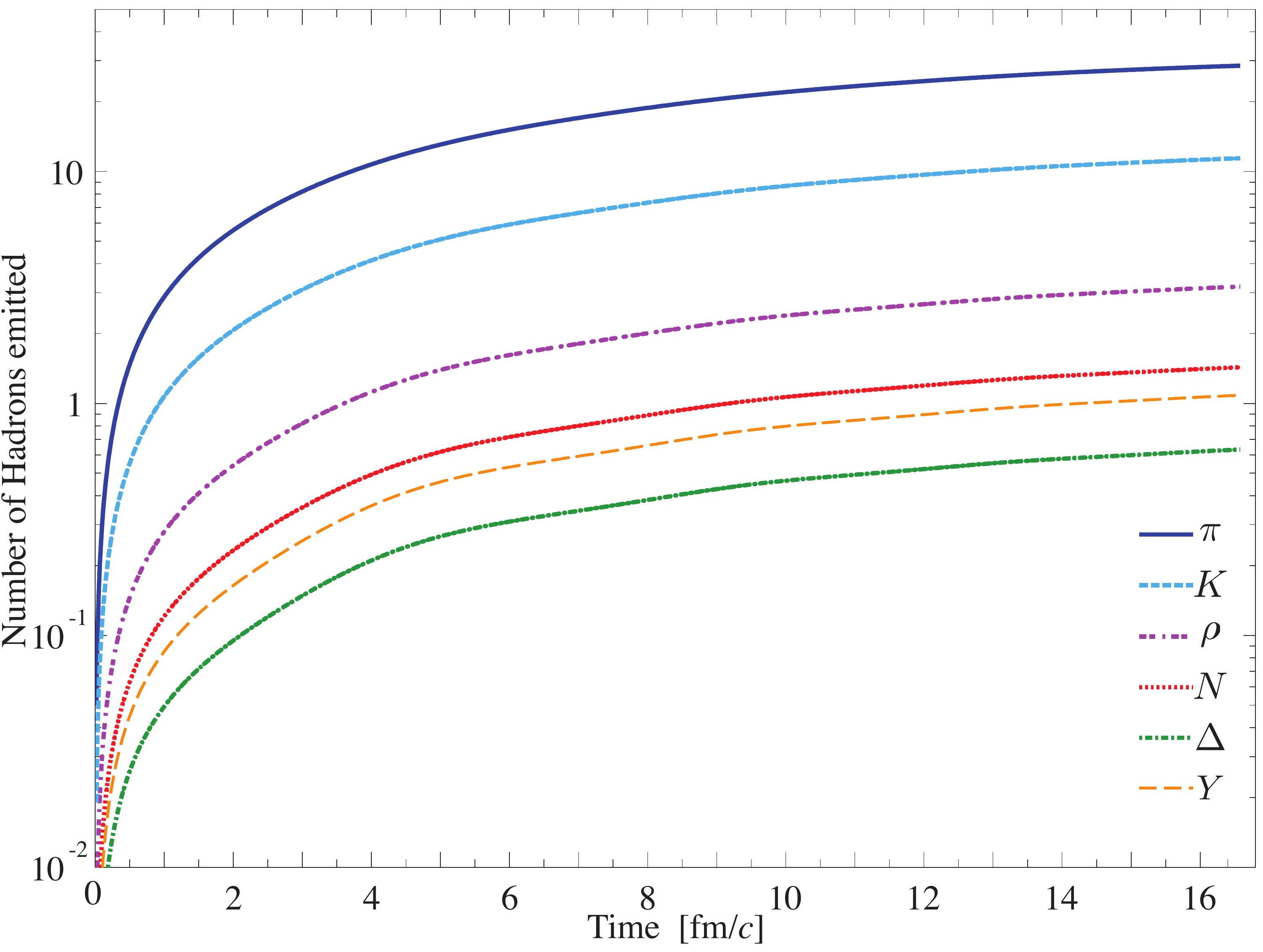}
\caption{\small Total numbers of specific hadrons (indicated in the figure) emitted as a function of time. The initial conditions are $T_0=154.3$ MeV, $R_0 = 2$ fm and $\dot{R}_0 = 0$.}\label{plot:ParticlesTime}
\end{figure}

Figures \ref{plot:Hadrons2}, \ref{plot:Energy2} and \ref{plot:ParticlesTime} show the hadron emission rates, energy loss rates and the number of hadrons emitted as functions of time for two different initial conditions as in Fig. \ref{plot:Energyplot}. While in Fig. \ref{plot:ParticlesTime}, only the zero-velocity case is shown. From Figs. \ref{plot:Hadrons2} and \ref{plot:ParticlesTime} one can see that, the pions and kaons, are by far the most abundant emitted particles, since the emission of heavier species is suppressed by their mass.  Nevertheless, the heavy particle spieces contribute significantly to the energy loss. It follows from Figs. \ref{plot:Hadrons2} and \ref{plot:Energy2}, that at initial stage the typical emission rates are about 3 pions, 1 kaon and 1 heavier hadron per fm/$c$. These particles carry away accordingly 1 GeV, 0.8 GeV and 0.8 GeV of energy per fm/$c$. From Fig. \ref{plot:ParticlesTime}, one can see that about 30 pions, 10 kaons and 7 heavier particles are emitted from the droplet with initial thermal energy of 30 GeV during its life time.

From this analysis, we can make two conclusions: First, the inclusion of heavier particles in the emission process leads to a significant enhancement of the energy emission rate, and second, even moderate initial collective velocity leads to strong droplet oscillations and pulsed hadron emission.

\section{Conclusion and outlook}
We have presented a simple hydro-kinetic model of a quark droplet evolution, which takes into account bulk and surface contributions to the energy, collective expansion and hadron emission from the surface. If hadron emission is disregarded, the droplet behavior is described by anharmonic oscillations. We have found an approximate solution for these oscillations using a super-ellipse parametrization of the phase-space trajectories. In the limit of small-amplitude, the period of oscillations is also proportional to the droplet radius. When hadron emission is included, the motion is changed to damped oscillations characterized by the decay time, which in first approximation is proportional to the droplet radius. For droplets with initial radii 1.5-2 fm these lines lie in the interval 9-16 fm/$c$. Our calculations show that pions account for about 2/3 of the emitted hadrons, but they carry away only 50 \% of the energy. 

In this paper we have considered only baryon-free droplet made of quarks and antiquarks with zero chemical potentials. This is a reasonable assumption for plasma produced at RHIC and LHC energies \cite{And11}. At lower collision energies effects of baryon asymmetry become more and more important. So, in the future, we are planning to include non-zero chemical potentials and investigate asymmetry between baryon and anti-baryon emission. Then, we can study such interesting phenomena as cold stranglet formation and flavor distillation \cite{Bar90,Gre87,Gre88}. The possibility of instantaneous hadronization of the dilute plasma \cite{Mis09} should be considered as well.

In our calculations, we have made an assumption that the droplet keeps a spherical shape at all times. This assumption can of course be questioned because the recoil effect from the emission, the Rayleigh-Taylor instability due to viscosity and collective quadrupole modes could lead to an asymmetric shape or even a bulged or fingered surface of the droplet. The recoil effect due to hadron emission can most likely be dismissed due to a stochastic nature of the emission process, i.e. the random directions of the emitted particles will lead to a cancellation of the recoil momenta due to individual particle emissions. Also, the surface tension should work against a non-spherical shape, trying to minimize the surface area of the droplet.

For a more realistic description, one could introduce viscous terms in the energy-momentum tensor, and study their impact on the dynamics of the droplet. It is clear, however, that these terms will damp collective oscillations but not change dramatically the life time of the droplets, which is mainly determined by the hadron emission rate. Also, the finite-size corrections to the droplet energy should be studied in more detail, in particular, higher-order terms in $1/R$ expansion should be introduced. These terms could play an important role at late states of the droplet evolution.

\section*{Acknowledgements}
The authors thank L.M. Satarov, G. Torrieri and S. Schramm for fruitful discussions. J.J.B.-B. acknowledge the fellowship received from the Helmholtz International Center for FAIR within the framework of the LOEWE program of the state of Hessen, Germany. I.N.M. acknowledges support provided by DFG grant 436RUS 113/7110-2 (Germany) and grant NS-215.2012.2 (Russia).

\end{document}